\documentclass[aps,prb,floatfix,superscriptaddress,noeprint]{revtex4-2}
\usepackage{amsmath,amssymb}
\usepackage{graphicx}
\usepackage[dvipsnames]{xcolor}
\usepackage{dsfont}
\newcommand*{\rom}[1]{\expandafter\@slowromancap\romannumeral #1@}

\newcommand{\bvec}{\boldsymbol}

\begin{document}
\title{Ultrafast Electron Dynamics in Altermagnetic Materials}

\author{Marius Weber}
        \affiliation{Department of Physics and Research Center OPTIMAS, University of Kaiserslautern-Landau, 67663 Kaiserslautern, Germany}
        \affiliation{Institut für Physik, Johannes Gutenberg University Mainz, 55099 Mainz, Germany}
\author{Kai Leckron}
\author{Luca Haag}
    \affiliation{Department of Physics and Research Center OPTIMAS, University of Kaiserslautern-Landau, 67663 Kaiserslautern, Germany}
\author{Rodrigo Jaeschke-Ubiergo} 
\author{Libor \v{S}mejkal}
\author{Jairo Sinova}

\affiliation{Institut für Physik, Johannes Gutenberg University Mainz, 55099 Mainz, Germany}
\author{Hans Christian Schneider}
    \affiliation{Department of Physics and Research Center OPTIMAS, University of Kaiserslautern-Landau, 67663 Kaiserslautern, Germany}

\date{\today}

\begin{abstract}
Altermagnets constitute a new class of magnetic materials that combine properties previously thought to be exclusive to either antiferromagnets or ferromagnets, and have unique properties of their own. In particular, a combination of symmetries connecting magnetic sublattices gives rise to a band spin splitting exhibiting unconventional d, g, or i-wave character.
 
Their unique electronic properties have already led to new spin-dependent transport effects. Here, we consider their spin and charge dynamics on ultrafast timescales. We use a minimal tight binding model that captures the main features of the altermagnetic candidate material KRu$_4$O$_8$. In the framework of this model, we compute the spin-dependent electronic scattering dynamics after ultrashort-pulse excitation and show through these microscopic calculations how electron-electron and electron-phonon scattering processes redistribute optically excited carriers in a 2D slice of the Brillouin zone.  
We find that the optically excited spin polarization is long lived ($\sim$1\;ps) compared to the electron-electron momentum scattering lifetime of roughly 10\;fs. This contrasts remarkably with the much shorter spin lifetimes observed in typical 
ultrafast electronic spin dynamics in conventional ferromagnets and antiferromagnets, making these pulse-driven spin excitation experiments a key probe of altermagnetism.
\end{abstract}

\maketitle

\section{introduction}
The exponentially increasing demand for storing and processing information in our AI digital age creates the need for pushing existing speed and efficiency limits ~\cite{jones2018information}. Present-day devices are  based on mature semiconductor and ferromagnetic thin-film technology~\cite{puebla_spintronic_2020}, so that further advances likely need new materials and/or tools for information manipulation. In order to take magnetoelectronics from the GHz frequencies of current ferromagnetic thin-film technology to the THz domain, antiferromagnetic spintronics have been proposed due to their intrinsically faster antiferromagnetic dynamics. However, compared to ferromagnets with their robust magnetoelectric effects, they are hampered by small achievable signals due to the weak spin-orbit coupling needed to manipulate the N\'{e}el vector~\cite{Smejkal2018,smejkal_route_2017,Nemec2018,Jungwirth2016}.
In order to overcome this roadblock, altermagnets have been introduced and classified as new class of magnetic materials~\cite{Smejkal2021a,Smejkal2022a,Smejkal2022GMR}. They constitute a possible material class for future non-silicon based information technology, as they combine the advantages of ferromagnets and antiferromagnets by exhibiting intrinsic THz dynamics with a robust, controllable magnetic splitting~\cite{gonzalez_betancourt_spontaneous_2023,feng_anomalous_2022,Bai2022,Liu2023,bai2024altermagnetism}. Altermagnets have so far been mostly studied for transport close to the Fermi energy and the intrinsic dynamics of their elementary excitations. In particular, novel effects such as the altermagnetic spin-splitter~\cite{Gonzalez-Hernandez2021} call for an investigation of altermagnets in the ultrafast time domain that can be accessed and probed by laser pulses. The present paper investigates the spin dependent electron dynamics on these ultrafast timescales in a d-wave altermagnet  including both electron-electron and electron-phonon scattering. By considering a planar d-wave altermagnet we can accurately describe the electronic dynamics using an effective model that captures the altermagnetic properties in a 2D slice of the Brillouin zone (BZ).  

\begin{figure}[t!]\centering\includegraphics[width=\linewidth]{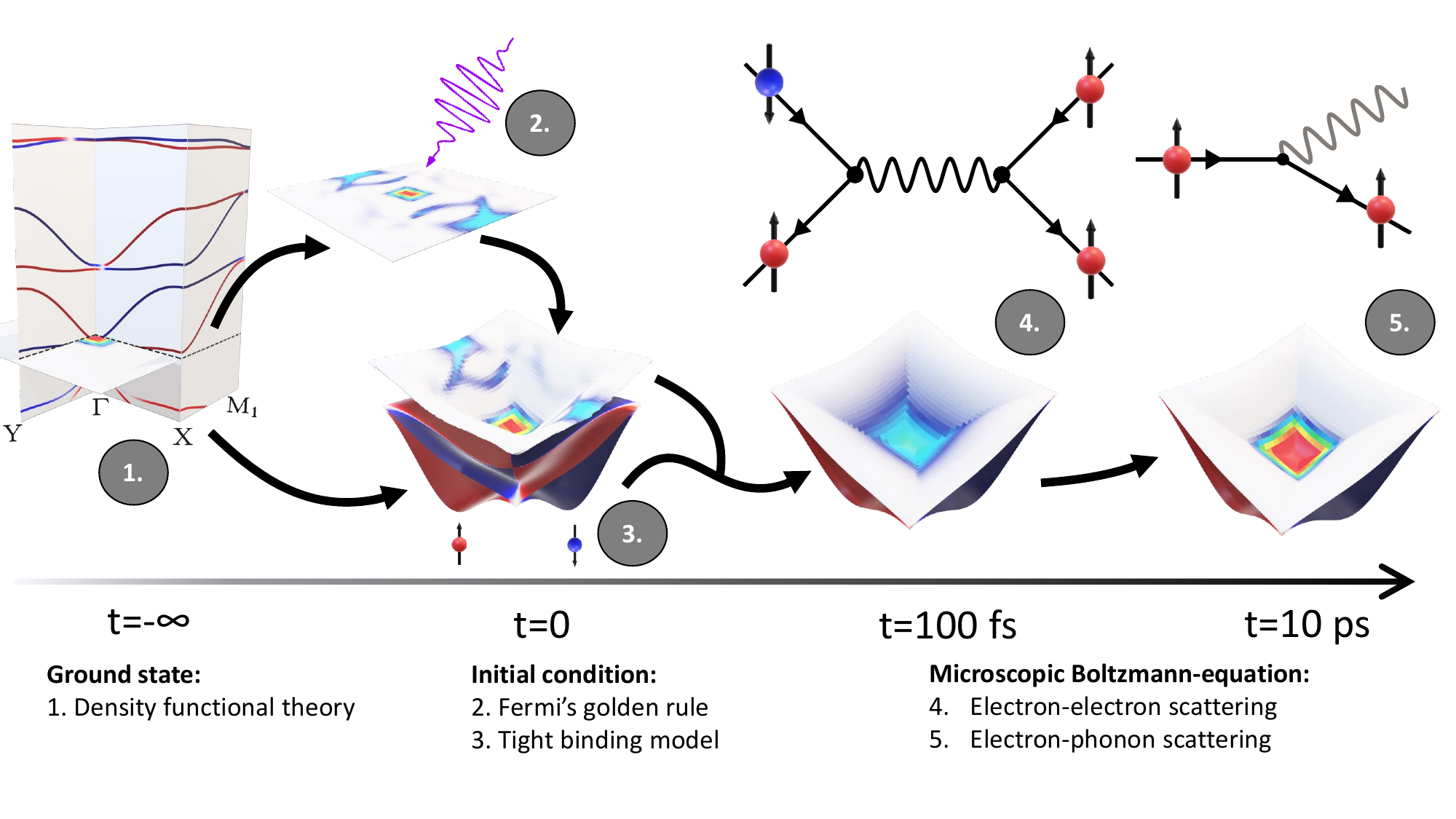}
    \caption{Sketch of altermagnetic properties and processes included:
    (1.) Dispersion of the altermagnetic bands along high symmetry paths (vertical planes) and electron distribution at initial equilibrium projected onto the $x$-$y$ plane at the Fermi-energy 
    (2.) Electron distribution after laser excitation 
    (3.) Altermagnetic conduction bands near the Fermi edge fit by tight-binding model 
    (4.) Optically excited electron-electron scattering dominates dynamics on a sub-ps time scale
    (5.) Electron-phonon scattering contributes to cooling/electronic energy dissipation on a sub-ns time scale.
    }
    \label{fig:f1}
\end{figure}

Figure~\ref{fig:f1} presents an overview of the steps involved in our approach to compute the electronic dynamics in an altermagnetic band structure (1.) Calculation of the band structure and the ground-state electronic distributions using density-functional theory (DFT); (2.) Computation of the laser-excited $k$-resolved electronic distributions; (3.) Fitting a minimal tight binding model to the two conduction bands close to the Fermi edge that are predominantly excited by the optical pulse, as calculated in step (2.). Mapping the excited electron distributions from the DFT to the corresponding tight-binding bands from step (2.); (4.) and (5.) Computing the electronic distribution functions in these bands including electron-electron and electron-phonon interaction processes at the level of Boltzmann scattering integrals. The last step is performed in the 2D slice of the Brillouin zone (BZ) in which the band structure is dominated by the planar d-wave character of the altermagnet. The reduction to the 2D slice of the BZ enables us to treat electron-electron scattering dynamics at the level of Boltzmann scattering integrals including the relevant $k$-dependent spin splitting in the band structure. A full 3-dimensional treatment of the scattering integrals will likely renormalize the results but not change the physics; details are presented in the Methods section.

\section{Band Structure and Optical Excitation \label{initial_condition}}

\begin{figure}[t!]
    \centering\includegraphics[width=\linewidth]{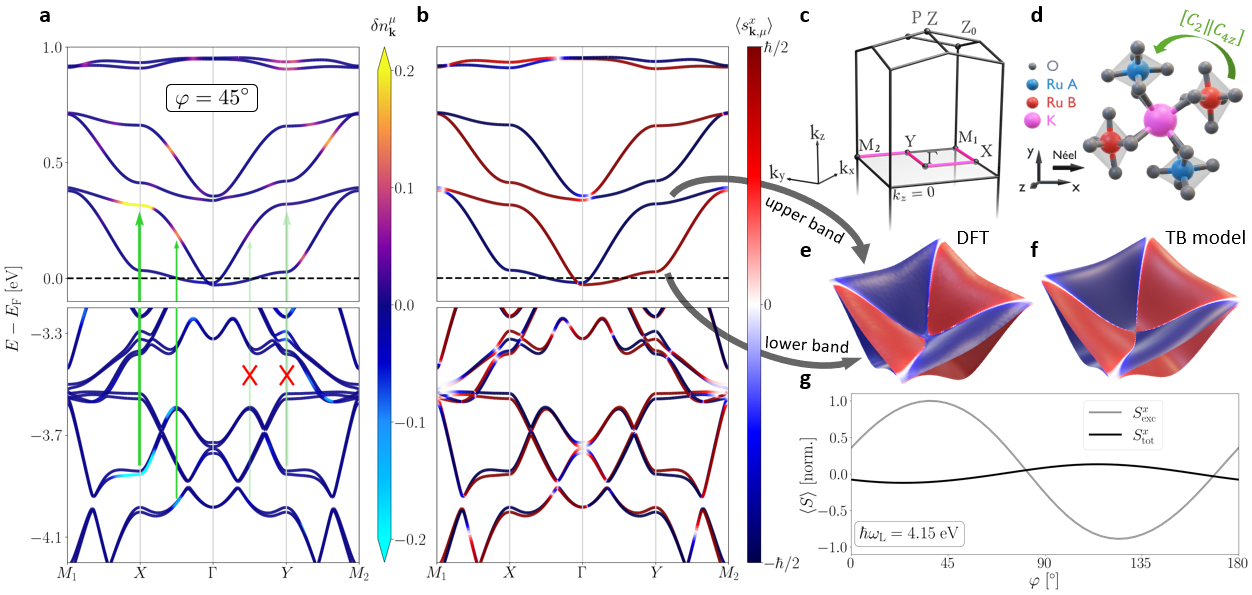}
    \caption{a) Band structure of KRu$_4$O$_8$ along the path $M_1-X-\Gamma-Y-M_2$; the color indicates the change in the distribution function due to the excitation with a linearly polarized laser pulse with a polarization vector which forms an angle of $45^\circ$ with the $[100]$ direction. b) Same band structure as in a), but colored according to the orientation of the spin expectation value. c) Sketch of the 3D Brillouin zone with the 2D Brillouin zone slice at $k_y=0$ highlighted as grey plane and a visualization of the path used in a) and c) as pink line. d) Real-space lattice structure of KRu$_4$O$_8$. e) Band structure and spin texture of the two bands close to the Fermi level according to DFT data. f) Band structure and spin texture of the minimal altermagnetic tight-binding model. g) Excited and total spin polarization for different orientations of the laser pulse polarization~$\varphi$.} 
    \label{fig:f2}
\end{figure}

We will first discuss the ground state properties of KRu$_4$O$_8$. We obtain quantities such as the band structure and the dipole matrix elements from \emph{ab initio} calculations~\cite{elk-code}. Figure~\ref{fig:f2}~b) shows the spin-resolved band structure along the high-symmetry path $M_1-X-\Gamma-Y-M_2$ in the BZ of KRu$_4$O$_8$. The $x$-component of the spin expectation values for the Bloch states at each $k$-point are shown in a color code where dark red and dark blue correspond to a complete spin polarization in $+x$ (``spin-up'') and $-x$ (``spin-down'') direction, respectively. In this color scheme, used in Fig.~\ref{fig:f1} and throughout the paper, white indicates a vanishing spin expectation value in $x$ direction. 

Figure~\ref{fig:f2}~c) illustrates the 2D slice of the Brillouin zone at $k_z=0$ that contains the path $M_1-X-\Gamma-Y-M_2$ by a shaded area and the path itself by a pink line. The d-wave character of the material is shown clearly in Fig.~\ref{fig:f2}~b) by the alternating spin orientation along $M_1-X-\Gamma-Y-M_2$ and in Fig.\ref{fig:f2} e) by the 2D spin texture, i.e., the map of $\langle s_k^x\rangle$ for the two bands around the Fermi edge. These two bands are energetically well separated and show particularly strong spin-mixing at the corners of the 2D BZ.

The excitation of the material by an ultrashort pulse is described by dipole transitions between single particle states $|\bvec{k}\mu\rangle$ corresponding to the Kohn-Sham energies~$\epsilon_{\bvec{k}}^{\mu}$. To this end we calculate the matrix elements $\mathbf{d}^{\mu\nu}_{\mathbf{k}}=\langle \mathbf{k}\mu|e \mathbf{r}|\mathbf{k}\nu\rangle$ of the dipole operator $\mathbf{d}=e\mathbf{r}$, where $e$ denotes the electronic charge. The distribution $n_{\mathbf{k}}^{\nu/\mu}$ of electrons in bands $\nu/\mu$ at momentum~$\bvec{k}$ created by the absorption of photons with energy $\hbar\omega_{\text{L}}$ are then obtained from
\begin{equation}
       \frac{\partial n_{\mathbf{k}}^{\mu}}{\partial t} \big\vert_{\mathrm{opt}}=\frac{2 \pi}{\hbar} \sum_{\nu}\big\vert d^{\mu\nu}_{\mathbf{k}} E(t) \big \vert^2 g(\epsilon^{\mu}_{\mathbf{k}}-\epsilon^{\nu}_{\mathbf{k}}-\hbar\omega_{\mathrm{L}})\big(n_{\mathbf{k}}^{\nu}-n_{\mathbf{k}}^{\mu}  \big) \,.  
       \label{optical}
\end{equation}
Here $E(t)$ is the magnitude of the electric field and $d=\mathbf{d}\cdot \hat{e}_\varphi$ is the projection of the dipole moment on the direction $\hat{e}_\varphi$ of the polarization vector of the E field. The spectral profile of the laser is contained in the broadening function $g(\epsilon^{\mu}_{\mathbf{k}}-\epsilon^{\nu}_{\mathbf{k}}-\hbar\omega_{\mathrm{L}})$.  

In this description of electronic dipole transitions, Eq.~\eqref{optical}, the optical pulse excites electrons from occupied states below the Fermi energy, i.e., $E_{\mathrm{F}}-\hbar\omega_{\mathrm{L}}$, to empty states above $E_{\mathrm{F}}$ according to the transition dipole matrix elements in this energy window. Fig.~\ref{fig:f2}~a) shows the resulting electron distribution for a polarization vector of the electric field $\hat{e}_\varphi$, i.e., rotated by $\varphi=45^{\circ}$ with respect to the [100] crystal direction. If we center the laser pulse around the photon energy $\hbar\omega_{\mathrm{L}}= 4.15$ eV, the dominant excitation appears in the two bands around the Fermi energy, where the highest excitation is close to $X$ in the upper band along the $\Gamma-X-M_1$  path, whereas only a weak excitation occurs along the  $\Gamma-Y-M_2$ path. Referring to Fig.~\ref{fig:f2}~b), we see that the region close to $X$ shows a pronounced spin-up character, such that this specific laser excitation leads to a electronic spin polarization in the conduction bands at the Fermi edge.

Figure~\ref{fig:f2}~g) shows a pronounced sinusoidal dependence of the strength of the excited spin polarization $\langle S^{x}\rangle_{\text{exc}} = \sum_{\boldsymbol{k},\mu,\epsilon_{\mathbf{k},\mu}\geq E_{\mathrm{F}}}n_{\boldsymbol{k},\mu} \langle s^x_{\boldsymbol{k}\mu}\rangle$ on the polarization angle of the laser $\varphi$. We denote here by $\langle s^x_{\boldsymbol{k}\mu}\rangle$ the $x$ component of the spin expectation value of the single-particle Bloch state $|\mu,\bvec{k}\rangle$. For instance, $\varphi= 45^{\circ}$ leads to a pronounced excitation in the spin-up channel whereas $\varphi=135^{\circ}$ leads to a similar result in the spin-down channel. It is also apparent from Fig.~\ref{fig:f2}~g) that the total spin polarization $\langle S^x \rangle_{tot} = \sum_{\boldsymbol{k},\mu}n_{\boldsymbol{k},\mu} \langle s^x_{\boldsymbol{k}\mu}\rangle$ is affected by the excitation process too, as the dipole-transition processes can change the spin expectation value between the initial and final Bloch states if these states are spin mixed due to spin-orbit coupling. For the excitation conditions used here, however, the change in total spin polarization is small.

The two bands around the Fermi edge that show the prominent altermagnetic characteristics in the ab-initio calculation are approximated by a minimal tight-binding model on a square lattice~\cite{Smejkal2021a} that captures the planar d-wave character of KRu$_4$O$_8$ and also includes spin-orbit coupling, see Supplementary Material. The corresponding eigenstates and thus the relevant interaction matrix elements can then be analytically obtained, which simplifies the following numerical calculation of the time evolution of the electronic system. The TB-Hamiltonian for the reciprocal lattice vector $\boldsymbol{k}:= a \boldsymbol{k}$ can be written in the form
\begin{equation}
    \mathcal{H}(\boldsymbol{k})=C(\boldsymbol{k})\sigma_0+ \Delta_c(\boldsymbol{k})\sigma_z+t^{I}_z(\boldsymbol{k})\sigma_x
    \label{eq:TBH}
\end{equation}
with nearest-neighbor hopping $C(\boldsymbol{k})=t_1 [\cos(k_x)+\cos(k_y)]+J$, the altermagnetic coupling terms, which involve the nearest-neighbor and second nearest-neighbor contributions $\Delta_c=t_2[\cos(k_x)-\cos(k_y)] +t_3\sin(k_x)\sin(k_y)$, and an energy shift $J$. Other parameters are listed in the Methods section~\ref{DFT_methods}. The nearest inter-sublattice hopping $t^{I}_z(\boldsymbol{k})=-4t_z \sin(\frac{k_y}{2})\sin(\frac{k_x}{2})$ accounts for spin-orbit coupling.
Table~\ref{table:1} summarizes the tight-binding parameters of the model band structure, which are determined by a fit to the DFT band structure of KRu$_4$O$_8$.

Figure~\ref{fig:f2}~f) shows the resulting TB band structure and spin texture obtained in the 2D BZ shown in Fig.~\ref{fig:f2}~c) as grey shaded plane. The eigenvectors of Eq.~\eqref{eq:TBH} are calculated analytically and the spin expectation values at each $k$-point are determined using the direction of the N\'eel vector ($x$) as quantization axis. The spin structure obtained from fitting the band structure is in good agreement with the \emph{ab initio} band structure and spin texture in Fig.~\ref{fig:f2}~e). 

\section{Electron dynamics \label{electron_dynamics}}
\begin{figure}[t!]
    \centering\includegraphics[width=\linewidth]{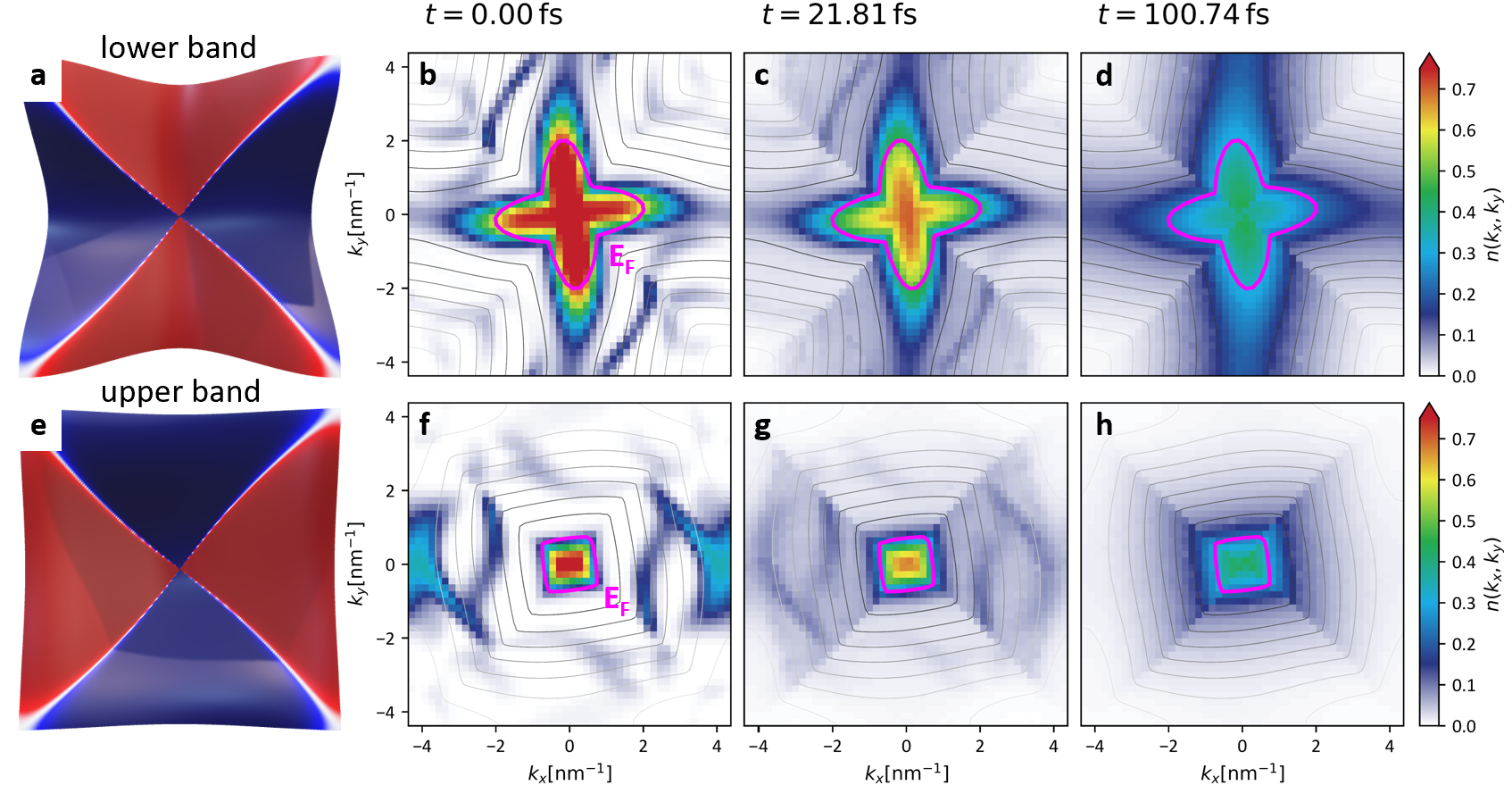}
    \caption{Spin textures of the optically excited bands a) lower and e) upper TB band from Fig.~\ref{fig:f2}~f). Snapshots of the electronic distribution functions $n_{\bvec{k}}^{\mu}$ in the 2D BZ for b)--d) lower and  f)--h) upper band for different times. The light grey to black lines indicate contour lines of the band structure, where the Fermi energy is highlighted in pink. Subplots b) and f) show the electron distributions directly after the instantaneous excitation, c) and~g) shortly after excitation, d) and~h) after momentum relaxation. Note that we use the unaltered data from the calculation on the $48\times 48$ $k$-grid and that we do not smooth out the distribution functions.}
    \label{fig:nk}
\end{figure}

The electronic dynamics, which are the main concern of this paper, are computed using an equation of motion (EOM) approach for the distribution functions $n_{\bvec{k}}^{\mu}$ of electrons in states $|\mu,\bvec{k}\rangle$. Here, the band index~$\mu$ implicitly contains the spin degree of freedom because the bands are spin split. For the time evolution of the electron distributions we solve the equation of motion~\cite{Haug-Koch,Rossi-Kuhn}
\begin{equation}
       \frac{\partial n_{\boldsymbol{k}}^{\mu}}{\partial t} \bigg|_{\mathrm{scat}}=\mathcal{S}^{\mu}_{\mathrm{e-e}}(\boldsymbol{k})+\mathcal{S}^{\mu}_{\mathrm{e-pn}}(\boldsymbol{k}).
       \label{eq:eom-tot}
\end{equation}
that treats the electron-electron and electron-phonon contributions~$\mathcal{S}^{\mu}_{\text{e-e}}$ and~$\mathcal{S}^{\mu}_{\text{e-pn}}$, respectively, at the level of Boltzmann scattering integrals. These are described in detail in the Methods section~\ref{electron_electron_methods} and~\ref{electron_phonon_methods}. 
The altermagnetic band and spin structure as well as the initial distribution of the excited electrons are decisively determined by the anisotropy in $k$-space, which therefore must be included in Eq.~(\ref{eq:eom-tot}) in order to correctly describe the spin-dependent electron dynamics.

We keep all the dependencies on the vector character of $\bvec{k}$ in the scattering kernels on a 2D BZ slice without introducing simplifying assumptions such as a density of states or an irreducible wedge~\cite{caruso, krauss_2009}. Our numerical solution assures that all the conservation laws, in particular the electron density and energy density, are fulfilled to a high degree of accuracy. Thus electron-electron scattering~\eqref{eq:eom-ee} cannot dissipate energy and can only lead to a ``hot'' Fermi-Dirac distribution by itself, even for long times.  Umklapp scattering processes are included in our calculation and thus the crystal momentum but not the total electron linear momentum is conserved. Further, electronic transitions between states with different spin mixing change the net spin polarization, so the total spin can be changed by the incoherent electron scattering processes included in Eq.~(\ref{eq:eom-tot}). Energy dissipation occurs due to electron-phonon scattering, which is included in (\ref{eq:eom-tot}) in a microscopic fashion with the phonon system treated as a bath, which acts as a momentum and energy sink. Therefore, in electron-phonon scattering events, only the electron density is conserved.

Due to the presence of SOC, electron-phonon scattering also effectively induces electronic Elliot-Yafet spin-flips~\cite{PhysRevLett.81.5624,PhysRevB.108.094403}. The excitation-induced increase in energy of the electrons is only dissipated by their scattering with phonons, which are assumed to always be at the equilibrium temperature of 150\,K. Since we are mostly interested on the ultrafast timescale $<10$\,ps and elevated excitation conditions where electron-phonon scattering is not the dominant contribution, we include only one longitudinal acoustic phonon branch.

Figure~\ref{fig:nk} displays the calculated electronic distribution functions for the carrier dynamics in the altermagnet after an excitation with an ultrafast laser pulse.  They illustrate the anisotropic character of the optical excitation in altermagnets, as electrons are mainly excited away from the $\Gamma$-point and favor one spin channel, here spin-up. Figs.~\ref{fig:nk}~a) and e) depict a top view of, respectively, the lower and upper band structure from Fig.~\ref{fig:f2}~f) with the spin-up contribution marked as red. The corresponding momentum-resolved distribution functions $n(k_x,k_y)$, are shown in Figs.~\ref{fig:nk}~b)--d) for the lower band and Figs.~\ref{fig:nk}~f)--h) for the upper bands. We include the energy dispersion as a contour line plot, where darker lines indicate lower energies and where the Fermi energy is highlighted as a pink line. Figures~\ref{fig:nk}~b) and~f) correspond to $t=0$, where we map the distributions as they are created by the optical pump field in the DFT band structure on the TB bands.

Figures~\ref{fig:nk}~c) and~g) present snapshots at roughly 20\,fs. Here, the electron distributions already exhibit a blurring of the sharp contours present in the initial electron distributions, which is a typical signature of the redistribution of electrons in the BZ by electron-electron scattering processes. Remarkably, this ``broadening'' of the excited electron distribution functions is contained within each spin-channel, indicating that spin-flip scattering does not play a major role on the 20 fs timescale. Figures~\ref{fig:nk}~d) and~h) show the corresponding distributions after 100\,fs. Here, the distinct excitation features in the spin-up channel have completely vanished as the excited carriers scattered towards a hot Fermi-Dirac distribution. Surprisingly, the imbalance between spin-up and spin-down channel, and thus the spin polarization, still persists on this timescale, which is especially well visible as the abrupt color switch at the diagonals, which form the borders of the spin-channels, in Fig.~\ref{fig:nk}~d). The anisotropic features in the electronic distributions persist long enough that they should be detectable in photoemission experiments~\cite{sven-aeschlimann}.

\begin{figure}[t!]
 \centering\includegraphics[width=\linewidth]{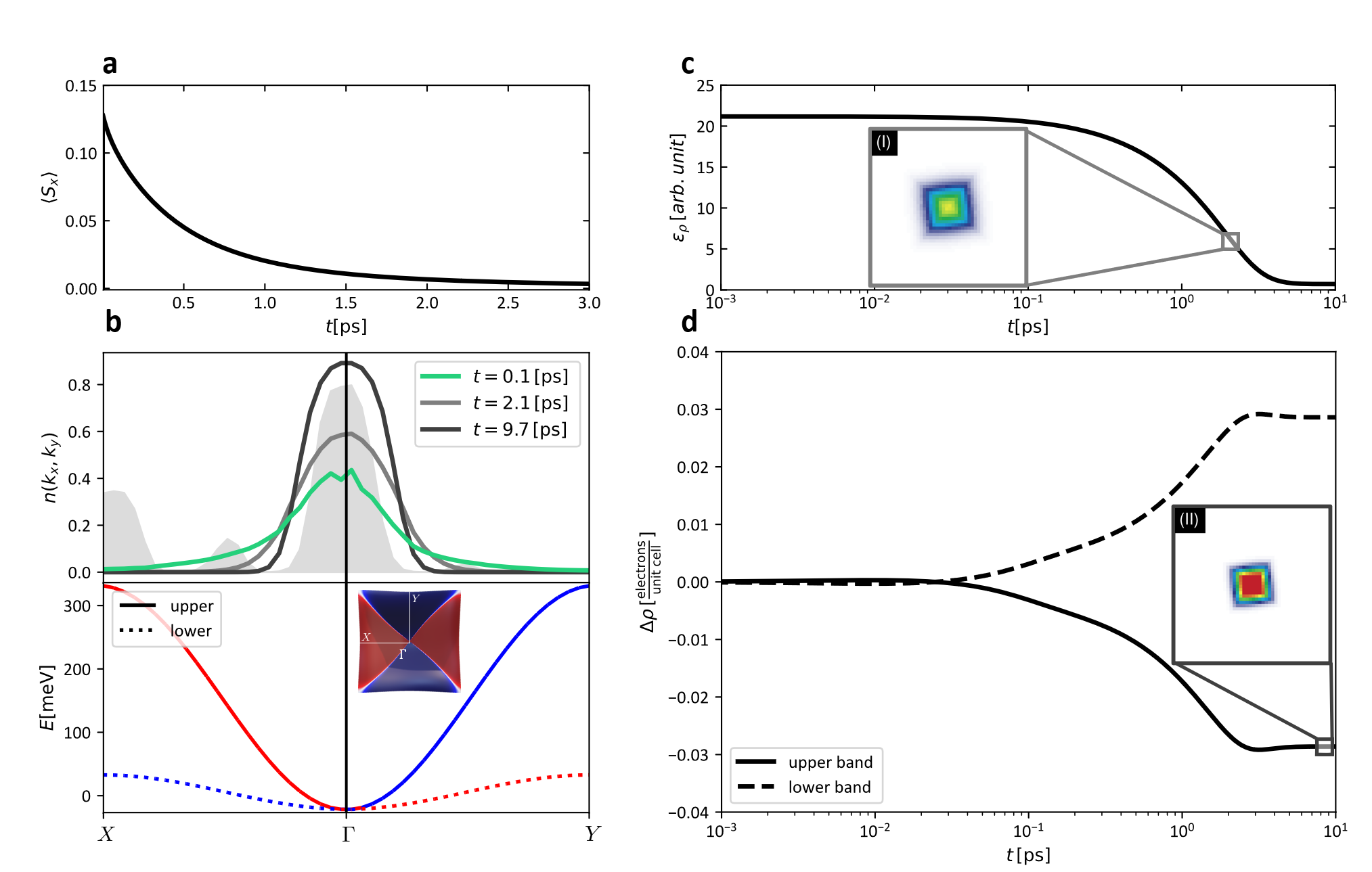}
    \caption{a) Dynamics of the ensemble spin expectation value in $x$-direction, b) top: distribution functions for selected times along the high symmetry path $X$--$\Gamma$--$Y$ where the shaded curve represents the initial distribution immediately after the pulse, bottom: corresponding spin-resolved band structure (inset shows the corresponding path in the BZ), c) energy density dynamics of the two considered bands and d) change of the band-resolved electron densities with respect to the densities of the excited electron distribution at t=0. The insets I and II  show the full 2D electron distribution at times 2.1 and 9.7 ps, respectively. Note that the time scales in c) and d) are logarithmic.}
    \label{fig:fig_4}
\end{figure}

Figure~\ref{fig:fig_4} characterizes the computed integrated spin and charge dynamics of the altermagnet.
The spin polarization in Fig.~\ref{fig:fig_4}a) decays from its initial value at $t=0$, which is determined by the the linearly polarized laser pulse with $\varphi = 45^{\circ}$ which has been chosen as the excitation condition for this calculation. It  drives electronic transitions between states with different spin mixing and thus creates the intial spin polarization that is subsequently affected by spin-changing intra and/or interband scattering processes. Underlying these spin dynamics are the carrier distributions shown in Fig.~\ref{fig:fig_4}~b), which are plotted along the high-symmetry path $X$--$\Gamma$--$Y$ in order to show how their anisotropic features develop over time. The corresponding band and spin structure along this path are displayed in the bottom panel.

The shaded area in the upper panel of Fig.~\ref{fig:fig_4}~b) represents the initial excited electron distribution with a sizable carrier density close to $X$ created by the excitation. After 100\;fs the most pronounced features at high energies have vanished due to electron-electron scattering. However, there is still a noticeable asymmetry between the electron distribution along the dominantly spin-up path $X$--$\Gamma$ and the distribution along the predominantly spin-down $\Gamma$--$Y$ path. Thus the spin polarization is almost unaffected on this timescale, cf. Fig.~\ref{fig:fig_4}~a), as the electron-electron scattering redistributes carriers largely in their own spin channels, i.e., inside the red or blue parts of the inset, which allows the spin polarization to persist on a significantly longer timescale than the intrinsic electron-electron scattering. After 2.1\,ps the scattering processes have led to distributions that are much closer to an isotropic form. At this time, most of the spin polarization has decayed. 

After 9.7\,ps the electron-phonon scattering has transferred most of energy added to the electronic system by the excitation process to the phonon bath. This is equivalent to a cooling of the electronic distribution functions, as indicated by the energy density plot in Fig.~\ref{fig:fig_4}~c).  
The energy is dissipated on a ps time scale, which points to the electron-phonon scattering as the relevant process. In turn, the ultrashort-time dynamics discussed in Fig.~\ref{fig:nk} are dominated by electron-electron scattering processes. 

Figure~\ref{fig:fig_4}~d) shows the band- resolved electron density change compared to the densities of the excited electron distributions in the two bands. Here, we can distinguish two regimes during which interband electron scattering processes occur.
Electron-electron scattering dominates within the first 100\,fs after which it becomes less efficient because most of its available scattering phase space has been reduced, cf.~Fig.~\ref{fig:nk}. The density change on the longer timescale between 0.5--10\,ps can be attributed to electron-phonon scattering, as can be concluded from the decay time of the energy density Fig.~\ref{fig:fig_4}~c). 
For even longer timescales, electron-hole recombination processes, which we do not include here, will return the system to complete equilibrium, i.e., electrons excited from below $E_{\mathrm{F}}$ to above $E_{\mathrm{F}}$ return. Figure~\ref{fig:fig_4} thus summarizes the different effects of the most important scattering processes on the altermagnetic spin polarization at the ultrafast timescale after an instantaneous excitation.

This long-lived spin polarization, as compared to the momentum scattering time, is a consequence of weak inter spin-channel scattering and is a rather robust effect. In particular, it does not depend sensitively on the material parameters, but only on the planar d-wave character of the band structure, and we expect this result to be valid for altermagnets with pronounced d-wave spin splitting in general. A long-lived spin polarization as shown in Fig.~\ref{fig:fig_4} should lead to a clear signal in time resolved magneto-optical measurements. Evidence of this unique altermagnetic feature has been found recently in experiments on ultrathin films of RuO$_2$~\cite{weber2024opticalexcitationspinpolarization}.

\section{Conclusion \label{discussion}}
Although the characterization of altermagnetic behavior has relied heavily on signatures in the angle-resolved photoemission spectroscopy (ARPES), ultrafast pulses combined with optical detection techniques promise to yield complementary insights on these materials~\cite{weber2024opticalexcitationspinpolarization}.
In order for optical techniques to become a characterization and manipulation tool as useful as in ferromagnets, one has to understand the optically driven carrier dynamics in such materials. We have presented a comprehensive theoretical study of the carrier dynamics of planar d-wave altermagnets on the femtosecond to picosecond timescale. A combination of \emph{ab initio} based calculation of electronic dipole transitions is used to calculate the momentum-resolved excitation conditions created by the absorption of a linearly polarized optical pump field in KRu$_4$O$_8$. Such an optical excitation creates anisotropic and spin-polarized carrier distributions, where the excited spin polarization depends on the polarization of the laser pulse. This connection between the \emph{linearly} polarized pump and the electronic spin polarization is a unique property of altermagnetic materials. The calculation of  $k$-resolved electronic distribution function dynamics due to electron-electron scattering and electron-phonon scattering in a 2D slice of the Brillouin zone for these realistic excitation conditions captures the inherently anisotropic carrier dynamics in altermagnets. Electron-electron scattering, which constitutes the fastest scattering process with a time scale of roughly 10 fs, predominantly equilibrates carriers in their own spin channel, leading to a surprisingly long-living spin polarization on the ps time scale. From the viewpoint of experiments~\cite{weber2024opticalexcitationspinpolarization}, it seems possible to measure the optically induced spin polarization with the time-resolved magneto optical Kerr effect (TR-MOKE) with a sufficiently good spectral resolution as the spin polarization lives longer than typical pulse durations. The calculations presented here also provide a theoretical basis for time-resolved angle and spin-resolved photoemission (TR-ARPES) experiments that can resolve the electron distribution in the entire Brillouin zone. Furthermore, it may even be possible that there are applications of the comparatively long-lived spin polarization in altermagnets that can be fine-tuned by optical pulses, including their polarization direction.

\begin{acknowledgments}
Funded by the Deutsche Forschungsgemeinschaft (DFG, German Research Foundation) 
– TRR 173 – 268565370 (projects A03, A08 and B03).
\end{acknowledgments}

\section{Methods}

\subsection{Density functional theory calculations \label{DFT_methods}}
The electronic single-particle properties are calculated using the full-potential linearized augmented plane wave code as implemented in the Elk Code within the local spin density approximation (LSDA)~\cite{PhysRevB.45.13244}. The lattice constants of KRu$_4$O$_8$ are set to $a=b=9.9456$\,\AA{} and $c=3.1673$ \AA{} which are comparable to the lattice constants in Refs.~\cite{kobayashi2009,djafri1985}. We perform a fixed spin-moment calculation on a $18\times18\times18$ $k$-grid. The magnetic moments of the Ruthenium atoms are chosen as $0.35$ $\mu_{\mathrm{B}}$ for the atoms located at sublattice A and $-0.35$ $\mu_{\mathrm{B}}$ for sublattice B, such that the N\'eel vector points in the [100] direction of the crystal. These values qualitatively match the band structure by \v{S}mejkal et al.~\cite{Smejkal2021a}.

The optically induced carrier dynamics are determined from Eq.~\eqref{optical} which describe transitions due to absorption and stimulated emission of photons. This process depends on the laser frequency $\omega_{\mathrm{L}}$ and the corresponding electric field amplitude $\mathbf{E}(t)$ together with the spectral profile $g(\epsilon)$. To account for the influence of the electric field polarization we write the the E-field vector as
\begin{equation}
    \mathbf{E}(t)=E_0\hat{\mathbf{e}}_{\varphi}e^{-4\ln(2)\frac{t^2}{\tau_{\mathrm{FWHM}}^2}}
\end{equation}
with the polarization vector $\hat{\mathbf{e}}_{\varphi}=(\cos\varphi,\sin\varphi,0)^T$ and the duration $\tau_{\mathrm{FWHM}}=40$ fs for the laser pulse. The spectral pulse profile is given by the Gaussian
\begin{equation}
    g(\varepsilon_{\mathbf{k}}^{\mu}-\varepsilon_{\mathbf{k}}^{\nu}-\hbar\omega_{\mathrm{L}})=\frac{1}{\sqrt{2\pi}}\frac{\sqrt{4\ln(2)}}{\Gamma}e^{-4\ln(2)\frac{\left(\vert\varepsilon_{\mathbf{k}}^{\mu}-\varepsilon_{\mathbf{k}}^{\nu}\vert-\hbar\omega_{\mathrm{L}}\right)^2}{\Gamma^2}}
\end{equation}
where $\Gamma=100$ meV describes the energetic smearing of the fs laser pulse.

The parameters for the fitted effective tight binding model are shown in table \ref{table:1}
\begin{center}
    \begin{table}[h!]
        \begin{tabular}{ |c|c|c|c|c| } 
             \hline
           $J$ & $t_1 $ & $t_2 $ & $t_3$& $t_z $ \\ 
             \hline
             182& -102.0 & 74.7 & -10 & -2.5 \\ 
             \hline
        \end{tabular}
    \caption{Fit parameters [$\mathrm{meV}$] for the analytical model}
    \label{table:1}
    \end{table}
\end{center}
We added an energy shift of $J=182\,[\mathrm{meV}]$ to align the corresponding band structure to the DFT results.

\subsection{Spin-dependent electronic dynamics\label{3d-2d}}

In metallic ferromagnets, many bands with strongly mixed orbital character cross the Fermi energy so that excited carriers usually find scattering partners somewhere in the band structure which makes their energy the most important quantity. It is in that case often appropriate \cite{caruso} to assume that electron dynamics in momentum space can be described by a random-$k$ approximation, i.e., by the spin-dependent density of states, or one can even apply phenomenological models such as the three temperature model~\cite{PhysRevLett.111.167204,mueller_relaxation_2013,roth_temperature_2012} which reduce the dynamics to the exchange of energy between electron system, spin system, and the lattice. The altermagnet presents a significant challenge for all of these simplifications because the band and spin structure are anisotropic, but still show the characteristic symmetry. This is important for the optical excitation, but also for the scattering processes, which can only be described by taking the momentum dependence into consideration. We achieve this for the planar altermagnet by including the full dependence of the electronic distributions on spin and carrier momenta in the important 2D slice of the BZ for the scattering contributions discussed next.

\subsection{Electron-electron scattering \label{electron_electron_methods}}
The electron-electron contribution to the carrier dynamics are at the level of Boltzmann scattering integrals and take the general form
\begin{widetext}
    \begin{equation}
        \begin{split}
       \mathcal{S}^{\mu}_{\text{e-e}}(\boldsymbol{k})&= \frac{2\pi}{\hbar} \sum_{\bvec{k}_{1} \bvec{l}}  \sum_{\mu_1\mu_2\mu_3} 
        (V_{\bvec{k}_1,\bvec{l}, \bvec{k},\bvec{l}+\bvec{k}_1-\bvec{k}}^{\mu_3\mu_2, \mu\mu_1})^2 \delta\left( \varepsilon_{\bvec{k}}^{\mu} + \varepsilon_{\bvec{l}+\bvec{k}_1-\bvec{k}}^{\mu_1} - \varepsilon_{\bvec{k}_1}^{\mu_3} - \varepsilon_{\bvec{l}}^{\mu_2} \right)\\
        &\hspace{2.4cm}\Big(\left(1-n_{\bvec{k}}^{\mu}\right)\left(1-n_{\bvec{l}+\bvec{k}_1-\bvec{k}}^{\mu_1}\right)n_{\bvec{l}}^{\mu_2}n_{\bvec{k}_1}^{\mu_3} - n_{\bvec{k}}^{\mu} n_{\bvec{l}+\bvec{k}_1-\bvec{k}}^{\mu_1} \left(1-n_{\bvec{l}}^{\mu_2}\right)\left(1-n_{\bvec{k}_1}^{\mu_3}\right)\Big) 
        \end{split}
        \label{eq:eom-ee}
    \end{equation}
\end{widetext}
with the Coulomb matrix element $V_{\bvec{k}_1,\bvec{l}, \bvec{k},\bvec{l}+\bvec{k}_1-\bvec{k}}^{\mu_3\mu_2, \mu\mu_1} =V_{q} \left\langle \bvec{k}_1,\mu_3|\bvec{k},\mu\right\rangle\left\langle \bvec{l},\mu_2|\bvec{l}+\bvec{k}_1-\bvec{k},\mu_1\right\rangle$. Here, $V_{\boldsymbol{q}}$ denotes the Coulomb potential, and the overlap of initial and final spin states determines the interaction matrix elements. The overlaps are dominated by a dependence on the spin degree of freedom:  For instance, spin-flip scattering is weaker than scattering within the same spin-channel due to the smaller overlap of the spin states. The delta function ensures energy conservation for electron-electron scattering transitions. 
Since the carrier distributions are assumed to be effectively independent of $k_z$ we reduce the calculation of the electronic dynamics to a 2D $k$-space, which formally corresponds to the $k_z=0$ slice of the BZ. If the carrier distributions are independent of $k_z$ in a region of thickness $|k_z|<\delta k$, one obtains an effective Coulomb potential for the 2D calculation. In the Supplement, it is shown that this potential is given by
\begin{equation}
            V_{q}^2 = \frac{e^4}{\mathcal{V}^2\varepsilon_0^2\varepsilon_b^2} \Bigg(
            \frac{\delta k_z}{\left(q_{\perp}^2+ \kappa^2\right)\left(q_{\perp}^2 + (\delta k_z)^2 + \kappa^2\right)} + \frac{\arctan\left(\frac{\delta k_z}{\sqrt{q_\perp^2+ \kappa^2}}\right)}{\left(q_\perp^2+ \kappa^2\right)^{\frac{3}{2}}}\Bigg),
\end{equation}
where $e$ is the elementary charge, $\mathcal{V}$ the normalization volume, $\varepsilon_0$ the vacuum permittivity, $\varepsilon_b$ the background permittivity which we set to 1. For the case of the planar altermagnet treated here, the region in $k_z$ encompasses almost the whole extent of the BZ in that direction~$k_{\text{max}}$ so that we take $\delta k_z = k_{\text{max}}=2\pi/a$. The momentum transfer in in the $(k_x, k_y)$-plane is denoted by  $q_\perp$ and we choose  $\kappa=22.7\;\mathrm{nm^{-1}}$ for the inverse screening length. The dynamical calculations were performed on a grid with $48\times 48$ $k$-points.

In order to ensure that the relevant conservation laws are obeyed by the numerical solution, we extended a numerical scheme that guarantees density conservation, which was originally developed for electron-phonon scattering~\cite{leckron2018} to electron-electron scattering. The conservation of the energy density in this approach is, while not exact, improved by orders of magnitude compared to a straightforward implementation. The small violation of energy conservation incurred in the electron-electron scattering dynamics becomes only noticeable on the timescale of several picoseconds, and on this timescale it will be suppressed by the cooling effect provided by electron-phonon scattering processes. 
The electron-electron scattering~\eqref{eq:eom-ee}, as implemented here, can therefore equilibrate the electronic distributions on sub-picoscond timescales to (hot) Fermi-Dirac distributions, but it cannot cool the electrons back to the equilibrium state before the excitation. 

\newpage

\subsection{Electron-phonon scattering \label{electron_phonon_methods}}

The scattering contribution due to electron-phonon interactions reads 
\begin{widetext}
\begin{equation}
	\begin{split}
		\mathcal{S}^{\mu}_{\text{e-pn}}(\boldsymbol{k}) = 
		\frac{2\pi}{\hbar} \sum_{\bvec{k}' \nu\lambda} \left|g_{\bvec{k}' \nu , \bvec{k} \mu}^{\lambda}\right|^2 \bigg[ 
		&\left( \left( 1+ N_{\bvec{k}' -\bvec{k}}^{\lambda} \right) \left(1 - n_{\bvec{k}}^{\mu} \right) n_{\bvec{k}'}^{\nu}  - N_{\bvec{k}'-\bvec{k}}^{\lambda} \left( 1 -n_{\bvec{k}'}^{\nu} \right) n_{\bvec{k}}^{\mu} \right)
		\delta \big(\Delta E_-^{\lambda}\big|_{\bvec{k}'\bvec{k}}^{\nu\mu} \big) \\
		- &\left(\left( 1+ N_{\bvec{k} -\bvec{k}'}^{\lambda} \right) \left( 1 - n_{\bvec{k}'}^{\nu} \right) n_{\bvec{k}}^{\mu}  - N_{\bvec{k}-\bvec{k}'}^{\lambda} \left( 1 -n_{\bvec{k}}^{\mu} \right) n_{\bvec{k}'}^{\nu} \right)
	    \delta \big(\Delta E_+^{\lambda}\big|_{\bvec{k}'\bvec{k}}^{\nu\mu} \big)  \bigg],
	\end{split}
	\label{eq:eom-epn}
\end{equation}
\end{widetext}
where the relevant energy differences for electronic transitions due to interaction with phonons of branch $\lambda$ are $\Delta E _{\pm}^{\lambda}\big|_{\bvec{k}'\bvec{k}}^{\nu\mu} := \varepsilon_{\bvec{k}}^{\mu} - \varepsilon_{\bvec{k}'}^{\nu} \pm \hbar \omega_{|\bvec{k}'-\bvec{k}|}^{\lambda}$ occur in the delta functions. Since we assume a phonon bath the phonon distribution functions $N_q^{\lambda}=n_B(\hbar\omega_q^{\lambda})$ are always given by a Bose-Einstein distribution~$n_B(\hbar\omega^{\lambda}_q)$ with the lattice temperature $T^{\mathrm{(eq)}} = 150\;\mathrm{K}$. 
We take into account one acoustic phonon branch and neglect the mode index $\lambda$ in the following. Its dispersion is given by $\hbar \omega_{q}= \hbar c_{\mathrm{pn}} \left|\sin\left(\frac{a}{2} q\right)\right|\frac{2}{a}$ with the lattice constant $a$ and speed of sound $c_{\mathrm{pn}} = 5 \, \mathrm{nm \; ps^{-1}}$. The electron-phonon matrix elements were used in the form~\cite{Essert_2011, leckron_ultrafast_2017}
\begin{equation}
    g_{\bvec{k}\mu, \bvec{k}_1\nu} \simeq \sqrt{ \frac{\hbar}{2m_{\mathrm{ion}}\omega_{q}}} D q \langle \bvec{k}\mu | \bvec{k}_1\nu \rangle,
    \label{eq:e-pn-matel}
\end{equation}
with $m_{\mathrm{ion}}=101.07 \, \mathrm{u}$, and we assumed a deformation potential $D=1$ eV~\cite{baral_re-examination_2016}. 
\bibliography{Ref,Group_Sinova}
\end{document}


\title{Supplement to \\
Ultrafast Electron Dynamics in Altermagnetic Materials}

\author{Marius Weber}
        \affiliation{Department of Physics and Research Center OPTIMAS, University of Kaiserslautern-Landau, 67663 Kaiserslautern, Germany}
        \affiliation{Institut für Physik, Johannes Gutenberg University Mainz, 55099 Mainz, Germany}
\author{Kai Leckron}
\author{Luca Haag}
    \affiliation{Department of Physics and Research Center OPTIMAS, University of Kaiserslautern-Landau, 67663 Kaiserslautern, Germany}
\author{Rodrigo Jaeschke-Ubiergo} 
\author{Libor \v{S}mejkal}
\author{Jairo Sinova}
    \affiliation{Institut für Physik, Johannes Gutenberg University Mainz, 55099 Mainz, Germany}
\author{Hans Christian Schneider}
    \affiliation{Department of Physics and Research Center OPTIMAS, University of Kaiserslautern-Landau, 67663 Kaiserslautern, Germany}

\date{\today}
\maketitle
\section{Electron Dynamics in the 2D Brillouin zone}
\begin{figure}
    \centering
    \includegraphics[width=\linewidth]{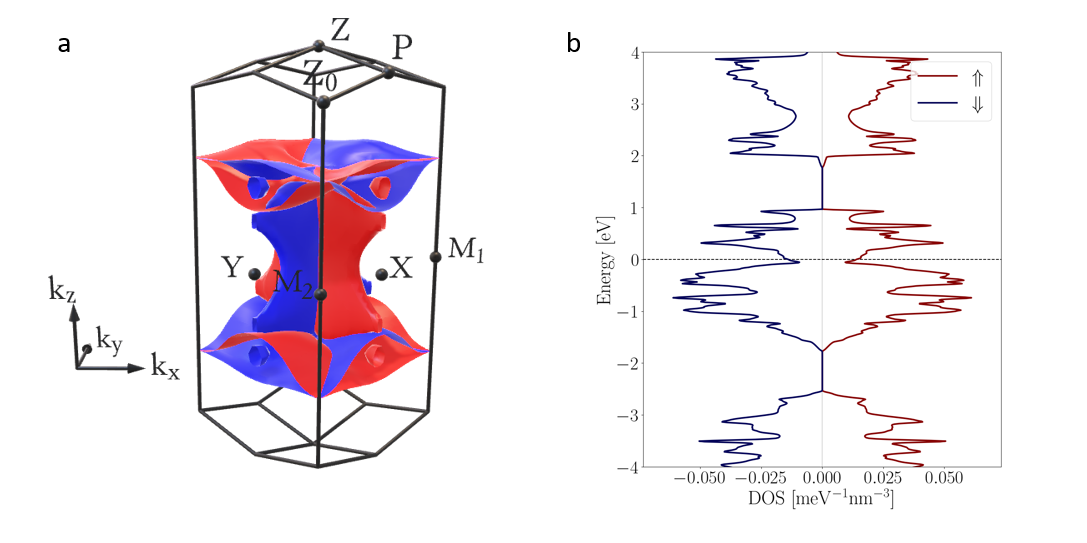}
    \caption{a) Spin-resolved Fermi surface and b) spin-resolved density of states (DOS) of KRu$_4$O$_8$. The Fermi surface is shown for a calculation without SOC to highlight the altermagnetic character, while the DOS is calculated including SOC. On the left (right) side in b) the density of states with mainly spin down (spin up) character is shown in blue (red). The Fermi energy is centered at zero (black dashed line). The densities on both sides are mirrored at the vertical line to highlight the compensated character of the altermagnet.}
    \label{fig:dos}
\end{figure}

We numerically calculate the electronic scattering dynamics by using a two-band model with an effective 2D Brillouin zone (BZ).
The reduction to a 2D BZ is suitable for a planar altermagnet such as KRu$_4$O$_8$, for which the spin resolved Fermi surface in the 3D BZ is depicted in Fig.~\ref{fig:dos}~a). The electronic states in this energy region show only a weak dependence on $k_z$ in a wide slice of the BZ. Neglecting the weak $k_z$ dependence in this volume yields an effective description of the energies and carrier distributions in terms of a 2D slice $(k_x,k_y)$ of the BZ alone. As the optical excitation does not introduce a pronounced $k_z$ dependence, the isotropic Coulomb interaction will not introduce it either. We can therefore determine the dynamics effectively for the reduced 2D Brillouin zone in $k_z=0$-plane by approximating $|(k_x,k_y,k_z),\nu\rangle\approx|(k_x,k_y,0),\nu\rangle$, and, in particular
\begin{equation}
    n_{(k_x,k_y,k_z)}^{\nu} \approx n_{(k_x,k_y,0)}^{\nu} \quad \text{and} \quad \varepsilon_{(k_x,k_y,k_z)}^{\nu}\approx \varepsilon_{(k_x,k_y,0)}^{\nu}.
    \label{eq:2d-n-epsilon}
\end{equation} 

Further, the evaluation of the dynamical equations including Coulomb scattering with $\bvec{k}$ vector resolution in the full Brillouin zone is not feasible at present. We next discuss the consequences of using the effective 2D BZ to describe the electron-electron Coulomb and electron-phonon scattering contributions to the equation of motion.

The general form of the electron-electron scattering contribution to  the equation of motion (EOM)~(6) is
\begin{equation}
        \begin{split}
       \mathcal{S}^{\mu}_{\text{e-e}}(\bvec{k}) &= \frac{2\pi}{\hbar} \sum_{\bvec{k}_{1} \bvec{l}}  \sum_{\mu_1\mu_2\mu_3} 
        (V_{\bvec{k}_1,\bvec{l}, \bvec{k},\bvec{l}+\bvec{k}_1-\bvec{k}}^{\mu_3\mu_2, \mu\mu_1})^2 \delta\left( \varepsilon_{\bvec{k}}^{\mu} + \varepsilon_{\bvec{l}+\bvec{k}_1-\bvec{k}}^{\mu_1} - \varepsilon_{\bvec{k}_1}^{\mu_3} - \varepsilon_{\bvec{l}}^{\mu_2} \right)\\
        &\hspace{2.4cm}\Big(\left(1-n_{\bvec{k}}^{\mu}\right)\left(1-n_{\bvec{l}+\bvec{k}_1-\bvec{k}}^{\mu_1}\right)n_{\bvec{l}}^{\mu_2}n_{\bvec{k}_1}^{\mu_3} - n_{\bvec{k}}^{\mu} n_{\bvec{l}+\bvec{k}_1-\bvec{k}}^{\mu_1} \left(1-n_{\bvec{l}}^{\mu_2}\right)\left(1-n_{\bvec{k}_1}^{\mu_3}\right)\Big) .
        \end{split}
        \label{eq:S-e-e-1}
\end{equation}
The dependence on the $z$-component of the momentum index here can be approximated as in Eq.~\ref{eq:2d-n-epsilon} for the distributions and electronic energies. Only the Coulomb matrix element depends on it in a more complicated way through the $q= |\bvec{k}_1-\bvec{k}|$ dependence of the  Coulomb potential $V_q$ because the overlap of the states is also $k_z$-independent.  
We thus consider the statically screened Coulomb potential in 3D momentum space, which  is given by $V_q^{3D} = \frac{e^2}{V\varepsilon_0\varepsilon_b\left(q^2+\kappa^2\right)}$ and replace the vector sums over $\bvec{k}_1$ and $\bvec{l}$ in Eq.~\ref{eq:S-e-e-1}. The resulting expression is
\begin{equation}
    \begin{split}
   \mathcal{S}^{\mu}_{\text{e-e}}(\bvec{k})  &= \frac{2\pi}{\hbar} \left(\frac{\mathcal{V}}{(2\pi)^3}\right)^{2}\int_{-k_\mathrm{max}}^{k_\mathrm{max}} \int_{-k_\mathrm{max}}^{k_\mathrm{max}} \int_{-\Delta k_z}^{\Delta k_z} \int_{-k_\mathrm{max}}^{k_\mathrm{max}} \int_{-k_\mathrm{max}}^{k_\mathrm{max}} \int_{-\Delta k_z}^{\Delta k_z}  
   \mathrm{d} l_z \; \mathrm{d} l_y \; \mathrm{d} l_x \; \mathrm{d} k_{1z} \; \mathrm{d} k_{1y} \; \mathrm{d} k_{1x} \\
   &\sum_{\mu_1\mu_2\mu_3} 
    (V_{\bvec{k}_1,\bvec{l}, \bvec{k},\bvec{l}+\bvec{k}_1-\bvec{k}}^{\mu_3\mu_2, \mu\mu_1})^2 \delta\left( \varepsilon_{\bvec{k}}^{\mu} + \varepsilon_{\bvec{l}+\bvec{k}_1-\bvec{k}}^{\mu_1} - \varepsilon_{\bvec{k}_1}^{\mu_3} - \varepsilon_{\bvec{l}}^{\mu_2} \right) \\
    &\Big(\left(1-n_{\bvec{k}}^{\mu}\right)\left(1-n_{\bvec{l}+\bvec{k}_1-\bvec{k}}^{\mu_1}\right)n_{\bvec{l}}^{\mu_2}n_{\bvec{k}_1}^{\mu_3} - n_{\bvec{k}}^{\mu} n_{\bvec{l}+\bvec{k}_1-\bvec{k}}^{\mu_1} \left(1-n_{\bvec{l}}^{\mu_2}\right)\left(1-n_{\bvec{k}_1}^{\mu_3}\right)\Big).
    \end{split}
\end{equation}
where $\Delta k_z$ denotes the width of the volume in $k_z$ direction for which we wish to obtain the effective description. For the resulting 2D slice, $k_{\mathrm{max}} = \pi/a$ describes its extent in $k_x$ and $k_y$ direction, using the lattice constants in $x(y)$ direction $a(b\equiv a)$. 
The Coulomb potential $V_q$ does not depend on $l_z$, since $q=|\bvec{k}_1-\bvec{k}|$,  so that the integral $\int \mathrm{d} l_z$ reduces to $2\;\Delta k_z$.
Finally, the integration over $k_{1z}$ can be evaluated analytically 
\begin{equation}
    \begin{split}
        \int_{-\Delta k_z}^{\Delta k_z} V_q^2 \; \mathrm{d} k_{1z} &= \int_{-\Delta k_z}^{\Delta k_z} \left(\frac{e^2}{\mathcal{V}\varepsilon_0\varepsilon_b\left((k_{1x}-k_x)^2+(k_{1y}-k_y)^2+(k_{1z}-k_z)^2 + \kappa^2\right)}\right) ^2 \mathrm{d} k_{1z} \\
        &= \frac{e^4}{\mathcal{V}^2\varepsilon_0^2\varepsilon_b^2} \Bigg(
        \frac{\Delta k_z}{\left(q_{\perp}^2+ \kappa^2\right)\left(q_{\perp}^2 + (\Delta k_z)^2 + \kappa^2\right)} + \frac{\arctan\left(\frac{\Delta k_z}{\sqrt{q_\perp^2+ \kappa^2}}\right)}{\left(q_\perp^2+ \kappa^2\right)^{\frac{3}{2}}}\Bigg) \\
    \end{split}
\end{equation}
For the screening parameter $\kappa$ we use the Thomas-Fermi result $\kappa=\sqrt{\frac{e^2}{\varepsilon_0}D(E_{\mathrm{F}})}$, where $D(E_\mathrm{F})$ denotes the density of states at the Fermi energy. Note that the normalization volumes $\mathcal{V}$ cancel out as they should with the other integrals over momenta. From the spin resolved density of states, as shown in Fig.~\ref{fig:dos}~b)  we obtain $\kappa=22.48$ nm$^{-1}$ for the screening parameter.

Electron-phonon scattering acts on a longer timescale than electron-electron scattering and is responsible for the energy dissipation to the lattice. Consequently, our treatment of the electronic coupling to phonons is simpler than electron-electron scattering. We use the long-wavelength limit for phonons in a 2D lattice
\begin{equation}
	g_{\bvec{k}\nu, \bvec{k}_1\nu_1} = \sqrt{\frac{\hbar^2}{2m_{\mathrm{Ion}}\mathcal{N}\hbar\omega_q}} D q \langle \bvec{k}\nu | \bvec{k}_1\nu_1\rangle,
    \label{eq:e-pn-matel}
\end{equation}
with the deformation potential $D$ and average ionic mass $m_{\mathrm{Ion}}$ of the atoms in the unit cell. In the dynamical calculation of the electronic distributions throughout the 2D Brillouin zone we also include umklapp scattering, so that we adapt the long-wavelength result for the whole BZ and include periodic boundary conditions.

\section{Tight-Binding Model}
For the calculation of the electron dynamics we employ a model band structure of a planar d-wave altermagnet including spin-orbit coupling. The model consists of a bipartite square lattice in 2D with a basis of 4 states, which can be labeled by orbital wave functions $\phi_A$, $\phi_B$ and spin states $\chi_{\pm}$. The nonrelativistic spin symmetries ($^24/^1m$)~\cite{Smejkal2021a} lead to   
\begin{equation}
    \mathcal{H}_0(\bvec{k})=t_1(\cos(k_x)+\cos(k_y))\chi_0 \sigma_0 +\Delta_c \chi_z \sigma_0+J \chi_z\sigma_z.
\end{equation}
Here,  $t_1$ is a normal hopping contribution and $\Delta_c=2t_2 (\cos(k_x)-\cos(k_y)))+2t_3 (\cos(k_x+k_y)-\cos(k_x-k_y))$ represents the altermagnetic nearest and next nearest hopping amplitudes. The constant exchange splitting is denoted by $J$.
The spin-orbit coupling contribution for this model is obtained by using the magnetic symmetry ($2\,'/m\,'$), following the approach of Attias \textit{et al.} \cite{attias2024intrinsic}. In the notation of that paper, we obtain
\begin{equation}
    \mathcal{H}_{\text{SOC}}(\bvec{k}) = \chi_x t^{R}_y(\boldsymbol{k}) \sigma_x+\chi_y t^{I}_z(\boldsymbol{k}) \sigma_y
\end{equation}
with $t^{R}_y(\boldsymbol{k})=4t_y\cos(\frac{k_y}{2})\cos(\frac{k_x}{2})$ and $t^{I}_z(\boldsymbol{k})=-4t_z \sin(\frac{k_y}{2})\sin(\frac{k_x}{2})$. In the notation of Ref.~\cite{attias2024intrinsic},  $t^R$ and $t^I$ refer to the real and imaginary parts of a complex hopping parameter. We neglect $t^{R}_y(\boldsymbol{k})$ in the following as it leads to spin mixing  around the $\Gamma$-point, which is absent in the DFT result.
Again as in Ref.~\cite{attias2024intrinsic}, we assume a comparatively large exchange contribution which effectively splits the four-dimensional model in energetically well separated two-dimensional subspaces. We choose 
$V_{+}=\{\phi_A(\boldsymbol{r})\chi_+, \phi_B(\boldsymbol{r}) \chi_{-} \}$ and  $V_{-}=\{\phi_A(\boldsymbol{r})\chi_{-}, \phi_B(\boldsymbol{r})\chi_{+}\}$ as basis for a reduced two-band model that is governed by the effective hamiltonian
\begin{equation}
\mathcal{H}_{\pm}(\boldsymbol{k})
=\begin{pmatrix} 
C(\boldsymbol{k}) \pm J+ \Delta_c(\boldsymbol{k})&  t^{R}_y(\boldsymbol{k})\pm t^I_z(\boldsymbol{k}) \\
  t^{R}_y(\boldsymbol{k}) \pm t^I_z(\boldsymbol{k}) &  C(\boldsymbol{k})\pm J -  \Delta_c(\boldsymbol{k})
\end{pmatrix}
\end{equation}
with $C(\boldsymbol{k})=t_1(\cos(k_x)+\cos(k_y))+J$ . 
Its eigenvalues are obtained as
\begin{align}
    E_{1,2}&=C(\boldsymbol{k})
    \pm \sqrt{\Delta_c^2+ (t^{R}_y)^2+(t^{I}_z)^2-2t^{R}_yt^{I}_z}.
\end{align}
with the corresponding eigenvectors 
\begin{equation}
    v_{1,2}= 
    \begin{pmatrix} \frac{(\Delta_c\mp \sqrt{\beta})}{t^{R}_y-t^{I}_z} \\ 1  \end{pmatrix}. 
    \label{eigenvectors}
\end{equation}
The vectors determine the overlap matrix elements in the scattering kernels. The parameters are chosen to approximate the DFT bandstructure of KRu$_4$O$_8$ well.
\bibliography{Ref.bib,Group_Sinova}